\documentclass[10pt,aps,twocolumn,nofootinbib]{revtex4-1}
\usepackage{amsfonts}
\usepackage{mathrsfs}
\usepackage{amsmath}
\usepackage{amssymb}
\usepackage[normalem]{ulem}
\usepackage{extarrows}
\usepackage{slashed}
\usepackage{isodateo}
\usepackage{graphicx}
\usepackage{xcolor}
\usepackage[bookmarksnumbered=true,bookmarksopen=true]{hyperref}
 \hypersetup{colorlinks,%
             linkcolor=[rgb]{0,0.3,0.6}, %
             citecolor=[rgb]{0,0.3,0.6}, %
             urlcolor=[rgb]{0,0.3,0.6}}%
\renewcommand{\rm}{\mathrm}
\begin{document}
\title{Identifying Minimal Composite Dark Matter}
\author{Shuai Xu}
\email{xushuai@zknu.edu.cn}
\affiliation{School of Physics and Telecommunications Engineering, Zhoukou Normal University, Henan 466001, China}
\author{Sibo Zheng}
\email{sibozheng.zju@gmail.com}
\affiliation{Department of Physics, Chongqing University, Chongqing 401331, China}

\begin{abstract}
We attempt to identify the minimal composite scalar dark matter from strong dynamics with the characteristic mass of order TeV scale.
We provide direct and indirect limits from dark matter direct detections and collider facilities.
Compared to a fundamental scalar dark matter, our results show that in the composite case with sizable derivative interaction between the dark matter and Higgs
the disappearing resonant mass region, the smaller spin-independent dark matter-nucleon scattering cross section in certain dark mass region,
and the absence at the HL-LHC provide us an opportunity to distinguish the composite dark matter.
\end{abstract}
\maketitle

\newpage
\section{Introduction}
Although a Higgs-like boson \cite{1207.7214,1207.7235} in the standard model has been established by the LHC,
there is still a lack of enough information about the ``nature'' of the observed Higgs.
Whether it is a fundamental or a composite state is critical as
it points to totally different new physics respect to the electroweak symmetry breaking.
This question will be addressed by near future precision measurements on the Higgs at HL-LHC \cite{1310.8361}.
Similarly,  there are also different choices on the thermal dark matter (DM),
which can be either fundamental or composite.
Since both the observed Higgs and yet confirmed DM are often simultaneously delivered by a single ``dark'' sector behind the electroweak symmetry breaking,
instead of conventional choice \cite{Xu:2019kuy}
in this paper we will explore both a composite Higgs and a composite scalar dark matter (CSDM).

In this scenario, the well-known hierarchy problem is solved by identifying both of these scalars as pseudo-Nambu-Goldstone (PNG) bosons \cite{composite1, composite2, composite3} tied to some global symmetry.
For reviews, see, e.g.\cite{1506.01961, 1005.4269}.
Following the spirit of simplicity, we consider the minimal CSDM with the following features.
\begin{itemize}
\item  The minimal structure of the coset which is suitable for both composite Higgs $h$ and CSDM $\eta$ is $\rm{SO(6)}/\rm{SO(5)}$ \cite{Gripaios:2009pe} based on the minimal composite Higgs model \cite{Agashe:2004rs}.
For recent discussions about alternative models, see e.g., \cite{Balkin:2017aep,Balkin:2018tma,Cacciapaglia:2019ixa}.
\item  The minimal matter content  in the effective theory of the composite sector contains only the light composite Higgs and CSDM, with the other freedoms therein decoupled.
\item The minimal representation of the composite fermions corresponds to the fundamental representation of $\rm{SO(6)}$.
\end{itemize}
The features above yield the following effective Lagrangian\footnote{Firstly, we use  Goldstone matrix to describe PNG bosons,
then derive kinetical terms in terms of Callan-Coleman-Wess-Zumino (CCWZ) formalism \cite{CWZ, CCWZ},
and finally calculate interactions between the PNG bosons and SM fermions as well as  effective potential in terms of spurion method (see, e.g.\cite{1506.01961}).}
for the PNG bosons in the minimal CSDM model
 \begin{eqnarray}{\label{L}}
\mathcal{L}_{\rm{eff}}&=&\frac{1}{2}\left(\partial_{\mu} h \right)^{2}-\frac{1}{2}m^{2}_{h}h^{2}-\frac{\lambda_{h^{3}}}{2}\upsilon h^{3}-\frac{\lambda_{h^{4}}}{4}h^{4} \nonumber\\
&+&\frac{1}{2}\left(\partial_{\mu} \eta \right)^{2}-\frac{1}{2}m^{2}_{\eta}\eta^{2}-\frac{\lambda_{\eta^{4}}}{4}\eta^{4}
+\mathcal{L}_{f}+\mathcal{L}_{V}-\mathcal{L}_{h}+\cdots,\nonumber\\
\end{eqnarray}
with
\begin{small}
\begin{eqnarray}
\mathcal{L}_{h} &=&\frac{\kappa_{1}}{2}\upsilon h\eta^{2} +\frac{\kappa_{2}}{4} h^{2}\eta^{2}
-\left(\partial_{\mu}h\partial^{\mu}\eta\right)\left[\frac{\xi}{\sqrt{1-\xi}}\frac{\eta}{\upsilon}+\frac{\xi(1+\xi)}{1-\xi}\frac{h\eta}{\upsilon^{2}}\right]\nonumber\\
&-&\frac{\xi^{2}}{1-\xi}\left(\partial_{\mu}h\right)^{2}\frac{\eta^{2}}{\upsilon^{2}},\label{L2}\\
\mathcal{L}_{f}&=&-m_{\psi}\bar{\psi}\psi\left[1+\frac{1-2\xi}{\sqrt{1-\xi}}\frac{h}{\upsilon}-\frac{(3-2\xi)\xi}{2(1-\xi)}\frac{h^{2}}{\upsilon^{2}}-\frac{\xi}{2(1-\xi)}\frac{\eta^{2}}{\upsilon^{2}}\right],\label{L3}\\
\mathcal{L}_{V}&=&(m^{2}_{W}W^{+}_{\mu}W^{-\mu} +\frac{m^{2}_{Z}}{2}Z_{\mu}Z^{\mu})\left[1+2\sqrt{1-\xi}\frac{h}{\upsilon}+(1-\xi)\frac{h^{2}}{\upsilon^{2}}\right],\nonumber\\ \label{L4}
\end{eqnarray}
\end{small}
where the constrained parametrization \cite{1204.2808, Marzocca:2014msa} has been adopted,
the Higgs mass $m_{h}=125$ GeV,  $m_\eta$ is the CSDM mass,
and $\xi=\upsilon^{2}/f^{2}$ with the weak scale $\upsilon=246$ GeV and $f$ referring to the breaking scale of $\rm{SO}(6)$.
We have neglected the next-to-leading-order terms, the derivative self interactions of the composite Higgs,
 and certain hidden parity such as $Z_2$ \cite{Gripaios:2009pe} with $\eta$ odd and the SM particles even
in order to ensure the stability of $\eta$.
Apart from the self interaction for $\eta$,
which is actually decoupled from both the DM relic abundance constraint and DM direction detections,
there are only three free parameters in Eqs.(\ref{L})-(\ref{L4}),
as composed of the CSDM mass $m_{\eta}$, the Higgs portal coupling
$\kappa_{1}=\kappa\sqrt{1-\xi}$ and $\kappa_{2}=\kappa(1-\xi)$,
and the composite mass scale $f$ or equivalently $\xi$,
which are responsible for phenomenologies of the minimal CSDM.
Although this observation is made in the constrained parametrization, it is also true in the other parameterizations, see e.g. \cite{Marzocca:2014msa}.

Instead of specific attempts in the literature,
the framework of parametrization as above enables us to estimate the general phenomenological status of the minimal CSDM.
In Sec.II, we will analyze the constraints on the CSDM from the latest DM direct detection limits.
Then, in Sec. III we turn to indirect constraints both from DM and collider experiments,
where the latest precision tests on the Higgs at the LHC are able to place strong indirect constraint on the parameter space.
Sec.IV is devoted to the direct probes of CSDM at the LHC.
During the study, we will point out the important differences between the CSDM and the fundamental scalar dark matter (FSDM) \cite{real1,real2} (for review, see e.g.\cite{1306.4710}).
We present the final results and conclude in Sec.V.

\section{DM Phenomenology}

\subsection{Parameter Space Of Dark Matter}
Instead of fixing $\kappa$ as in ref.\cite{Marzocca:2014msa},
which results from certain specific assumptions on the composite fermions in the composite sector,
we take it as a free parameter with rational values.
A relaxation on the parameter $\kappa$ gives rise to a parameter space of thermal DM obviously larger than that in ref.\cite{Marzocca:2014msa}.

\begin{figure}
\centering
\includegraphics[width=9.5cm,height=9cm]{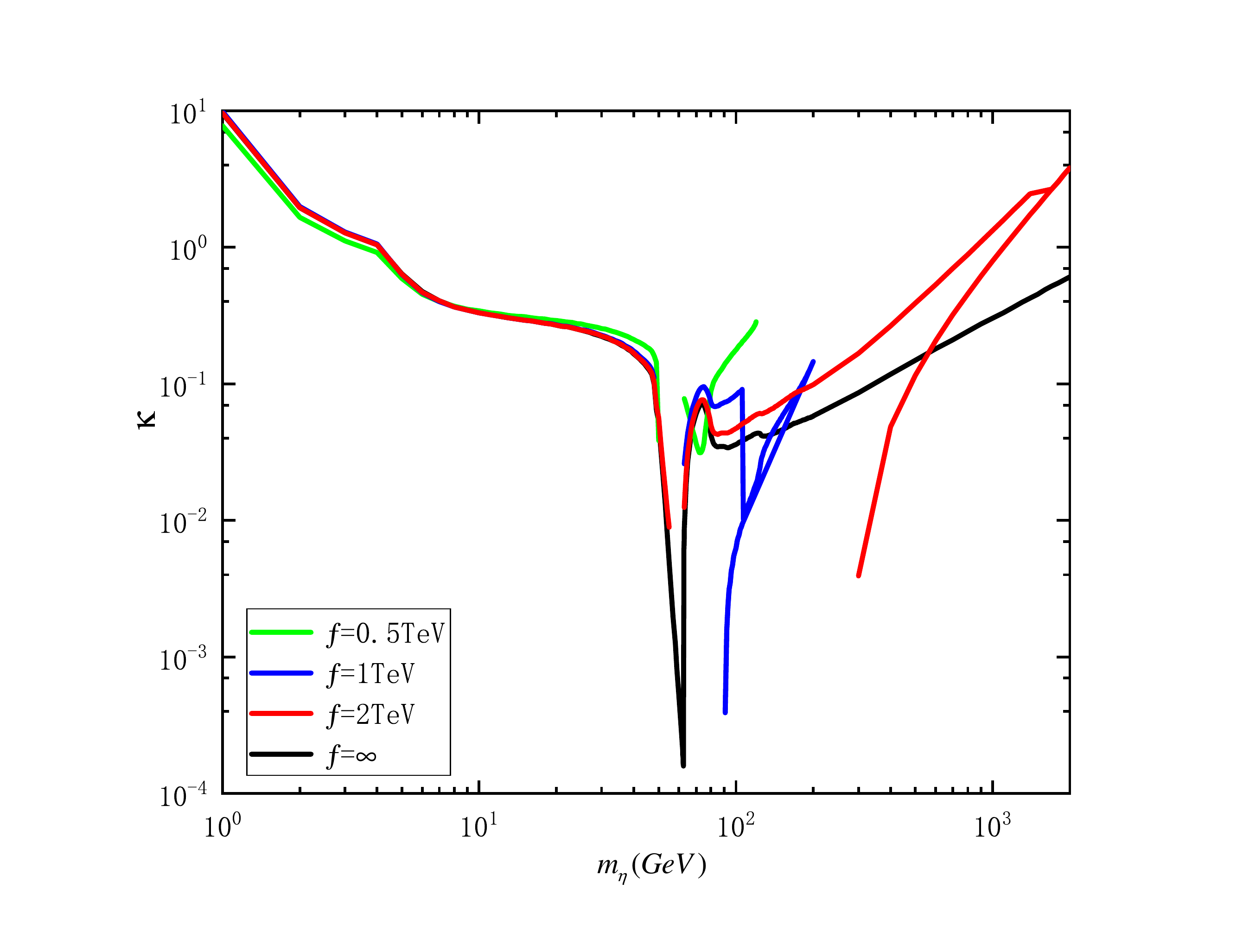}
\centering
\caption{Contours of thermal DM relic density projected to the two-parameter plane of $m_{\eta}-\kappa$ for the representative scales $f/\rm{TeV}=0.5$ (green), $1$ (blue) and $2$ (red),
where the black curve refers to the FSDM with $f=\infty$.
The dips near 60 GeV are due to the Higgs resonance, and the others occur wherever the derivative and non-derivative term in Eq.(6) nearly cancel each other. See text for details about the two-value phenomena.}
\label{relic}
\end{figure}

Apart from a partial of interactions similar to that of FSDM \cite{1306.4710, Han:2015hda} in Eq.(\ref{L2}),
$\mathcal{L}_{h}$ also contains  the derivative interactions with momentum dependence,
which lead to significant deviation from the FSDM for $f$ of order TeV scale.
The derivative interactions contribute to $\eta\eta$ annihilation cross section in the manner that it grows as the DM mass increases,
which can be interpreted from the modifications to the effective couplings in Eq.(\ref{L2}):
\begin{eqnarray}{\label{shift}}
\kappa_{1}&\approx& \kappa- 2\xi \left(\frac{m^{2}_{\eta}}{\upsilon^{2}}\right)+\mathcal{O}(\xi^{2}),\nonumber\\
\kappa_{2}&\approx&  \kappa- 8\xi \left(\frac{m^{2}_{\eta}}{\upsilon^{2}}\right)+\mathcal{O}(\xi^{2}).
\end{eqnarray}
In Eq.(\ref{shift}), the deviations are small in the limit $f\rightarrow \infty$,
which corresponds to the FSDM as shown by the black curve in Fig.\ref{relic}.
In contrast, the deviation is expected to be large in the case of small $f$,
under which $\kappa$ shifts from the value $\kappa_{\star}$ \cite{1306.4710, Han:2015hda} referring to the FSDM as
\begin{eqnarray}{\label{shift2}}
\left|\kappa-\left(\frac{2m^{2}_{\eta}}{\upsilon^{2}}\right)\xi \right|\approx  \kappa_{\star}.
\end{eqnarray}

Fig.\ref{relic} shows the parameter space respect to the thermal DM relic abundance $\Omega\rm{h}^{2}=0.12\pm0.001$ \cite{Aghanim:2018eyx} in terms of micrOMEGAs \cite{1407.6129}.
Compared to the FSDM, two-value phenomena appears as a result of the derivative interaction between $\eta$ and $h$. One can seek a hint as follows.
 If we switch off the derivative interaction, the two-value phenomena disappears.
 There would be only an overall correction on the magnitude of $\kappa_{\star}$.
 Instead, if we turn on the derivative interaction, it contributes to the vertex of the cubic interaction $\eta-\eta-h$ by a linear shift in the value of $\kappa$ in Eq.(\ref{shift}) in momentum space.
 Since the derivative interaction is essential in this model,
 the two-solution phenomena is inevitable when both of solutions in Eq.(\ref{shift2}) are positive and rational,
 which occurs in the parameter region with both large $\xi$ and $m_{\eta}$
 where the effect of derivative interaction is sizable. Instead,
in the parameter region with either small $m_{\eta}$ or small $\xi$ where the effect of derivation interaction is weak,
only single value of $\kappa$ is allowed as shown in Fig.\ref{relic}.
\emph{Due to the derivative interaction,
the well-known resonant mass window $m_{\eta} \sim m_{h}/2$ gradually disappears as $f$ approaches to smaller value}.

In the discussion above, we have neglected the contribution to the DM annihilation cross section from the contact interaction in Eq.(\ref{L3}),
which is given by,
\begin{eqnarray}{\label{shift3}}
\sigma(\eta\eta\rightarrow \bar{\psi}\psi)v_{\rm{rel}}\approx\frac{\xi^{2}}{16\pi}\frac{m^{2}_{\psi}}{\upsilon^{4}}\left(1-\frac{m^{2}_{\psi}}{m^{2}_{\eta}}\right)^{3/2}.
\end{eqnarray}
Because of the fermion mass suppression in Eq.(\ref{shift3}), the contribution can be indeed ignored in the CSDM mass range $m_{\eta}<m_{t}$.
Even for the CSDM mass range $m_{\eta}> m_{t}$ as covered by the case with $f=2$ TeV in Fig.\ref{relic},
$\sigma(\eta\eta\rightarrow \bar{t}t)v_{\rm{rel}}$ is small compared to the inferred value $\langle\sigma v_{\rm{rel}}\rangle\approx 3\times 10^{-26}\rm{cm}^{3}\rm{s}^{-1}$,
which indicates that the previous estimate on the behavior of this curve is still valid.

\begin{figure}
\centering
\includegraphics[width=9.5cm,height=9cm]{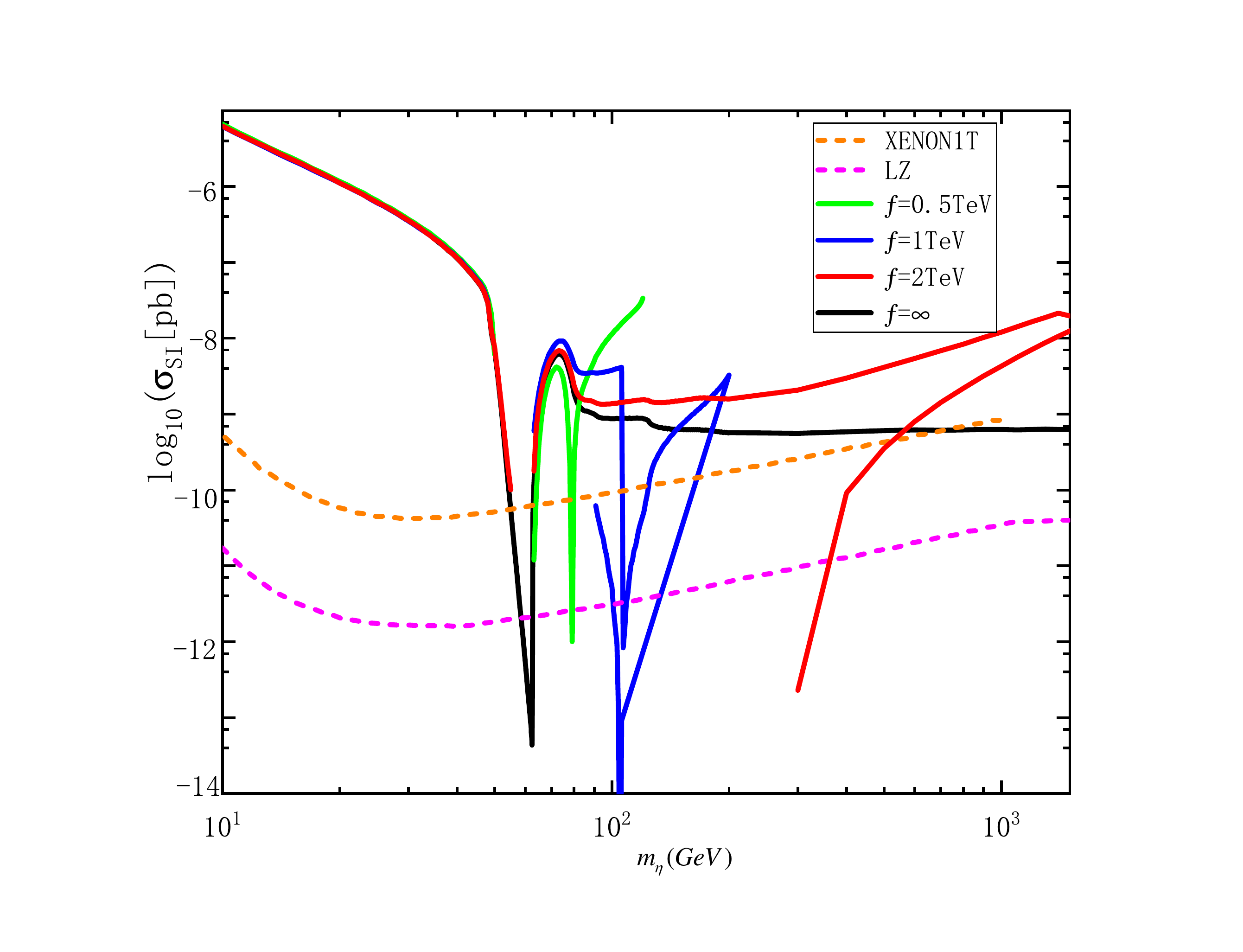}
\centering
\caption{SI cross section as function of $m_{\eta}$ for the representative scales as in Fig.\ref{relic},
which reveals that in certain CSDM mass range between $63$ GeV and $600$ GeV an improved XENON1T or future LZ results are useful in identifying the CSDM from the FSDM.}
\label{SI}
\end{figure}

\subsection{Direct Detection}
Combing the Higgs-portal interaction in Eq.(\ref{L2}) and the contact interaction in Eq.(\ref{L3})
yield the spin-independent (SI) scattering cross section for the CSDM
\begin{eqnarray}{\label{directcs}}
\sigma_{\text{SI}}=\frac{ f^{2}_{N}}{4\pi}\frac{\mu^{2}m^{2}_{N}}{m^{2}_{\eta}}\left(\frac{\kappa(1-2\xi)}{m^{2}_{h}}+\frac{\xi}{(1-\xi)\upsilon^{2}}\right)^{2},
\end{eqnarray}
where $m_{N}$ is the nucleon mass,
$\mu=m_{\eta}m_{N}/(m_{\eta}+m_{N})$ is the DM-nucleon reduced mass,
and $f_{N}\approx 0.3$ is the hadron matrix element.
Unlike the preceding analysis on the DM relic abundance,
the corrections to Eq.(\ref{directcs}) due to the derivative interactions are negligible.

Fig.\ref{SI} explicitly shows  the numerical results about the SI cross sections extracted from the DM parameter space in Fig.\ref{relic}.
In this figure, one finds that unlike the FSDM,
in which DM mass below $\sim 700$ GeV
is nearly excluded by the latest XENON1T limit \cite{Xenon1T},
a large part of the CSDM mass window between $\sim 63- 600$ GeV is still beneath the latest XENON1T limit \cite{Xenon1T} and can be in the reach of future LZ experiment \cite{LZ}.
For example,
the critical bound for $f=500$ GeV has been altered from $m_{\eta}\sim 70$ GeV by XENON100 limit \cite{1204.2808} to be nearly excluded by the XENON1T limit,
whereas the critical bound for $f=1$ TeV has been changed from $\sim 200$ GeV by LUX 2013 limit \cite{Marzocca:2014msa} to $\sim 150$ GeV by the XENON1T limit.
Therefore, \emph{the future XENON1T or LZ results in the large mass window can be useful in distinguishing the CSDM from the FSDM}.

\begin{figure*}
\centering
\includegraphics[width=8cm,height=9cm]{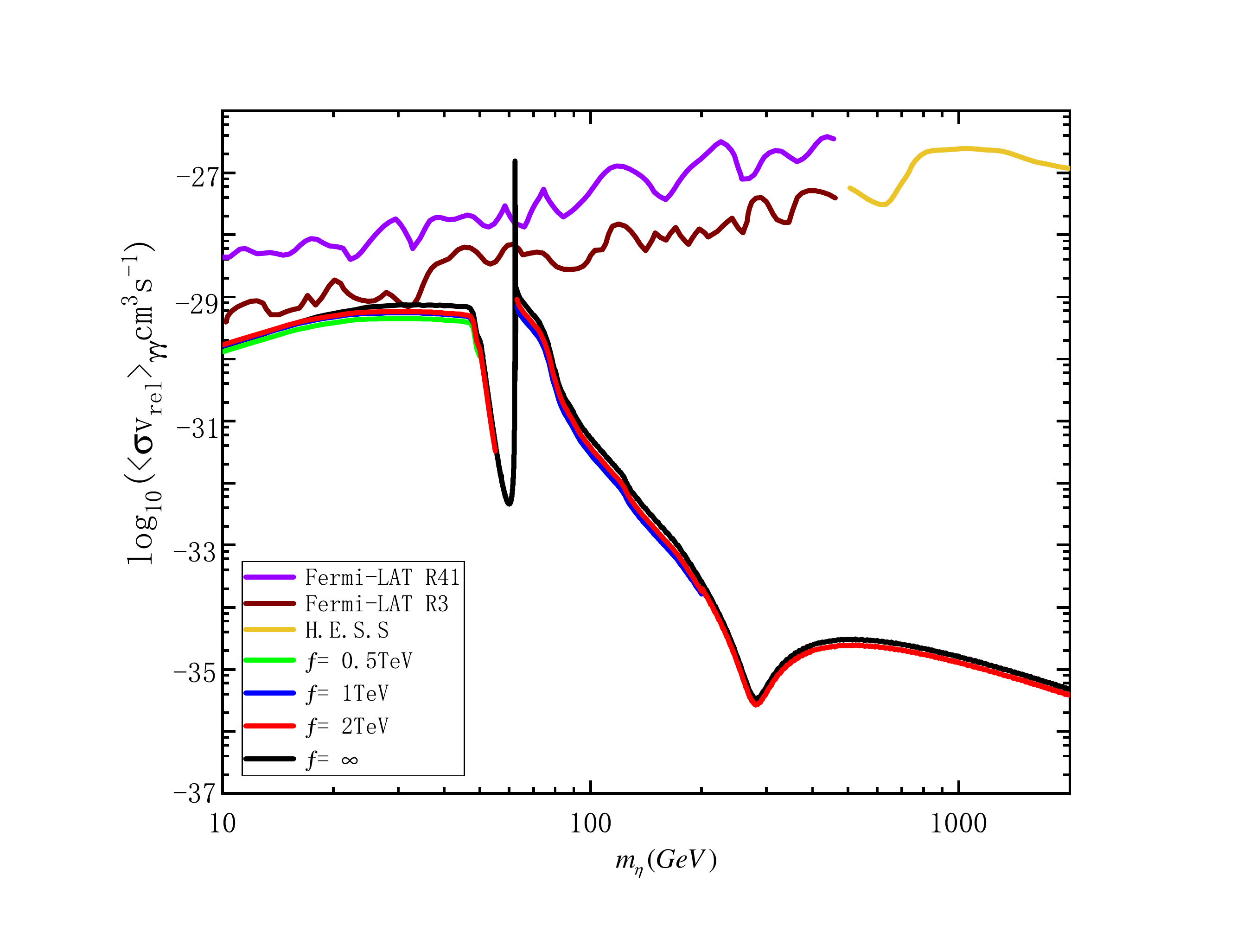}
\includegraphics[width=8cm,height=9cm]{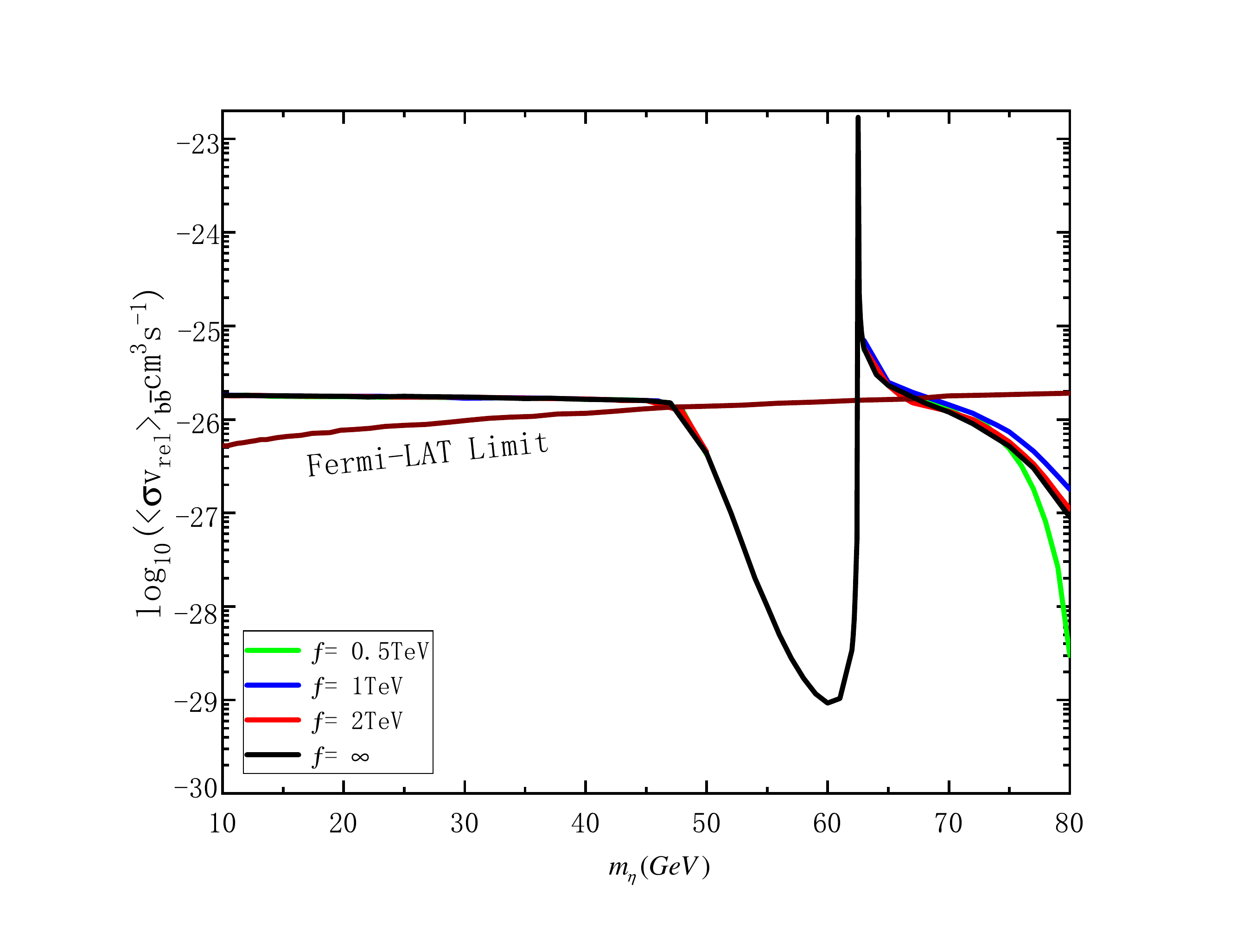}
\centering
\caption{The averaged cross sections of $\langle\sigma_{\gamma\gamma} \upsilon_{\rm{rel}}\rangle$ (left) and $\langle\sigma_{b\bar{b}} \upsilon_{\rm{rel}}\rangle$ (right) for various scales $f$ as in Fig.\ref{relic}.
While the $\gamma$ ray is weak, the $b\bar{b}$ limit from Fermi-LAT excludes CSDM mass regions $m_{\eta}\leq 47$ GeV and $63$ GeV $\leq m_{\eta}\leq$ $67$ GeV.}
\label{gammaray}
\end{figure*}

\section{Indirect Constraints}
\subsection{DM Annihilation}
Astrophysical observations from DM annihilation can be used to indirectly constrain the thermal DM.
For DM annihilation into $\gamma$ ray  the cross section can be calculated via standard formula \cite{Average},
\begin{eqnarray}{\label{average}}
\langle\sigma_{\gamma\gamma} \upsilon_{\rm{rel}}\rangle&=&\frac{x}{16m_{\eta}^{5}K^{2}_{2}(x)}\nonumber\\
&\times& \int^{\infty}_{4m^{2}_{\eta}} ds \sqrt{s-4m^{2}_{\eta}}sK_{1}\left(x\sqrt{s}/m_{\eta}\right)\sigma_{\gamma\gamma} \upsilon_{\text{rel}}, \nonumber\\
\end{eqnarray}
where $x=m_{\eta}/T$, $s$ is the square of the center-of-mass energy,
and $K_1$ and $K_2$ are modified Bessel functions of the second kind.
In Eq.(\ref{average}), the annihilation cross section reads as,
\begin{eqnarray}{\label{gammacs}}
\sigma_{\gamma\gamma} \upsilon_{\text{rel}}=\frac{2\upsilon^{2}}{\sqrt{s}}\left(\kappa-\frac{2m^{2}_{\eta}}{\upsilon^{2}}\xi\right)^{2}\frac{\Gamma_{h\rightarrow\gamma\gamma}}{(s-m^{2}_{h})^{2}+m^{2}_{h}\Gamma^{2}_{h}},
\end{eqnarray}
where $\Gamma_{h}\approx 4.15$ MeV is the total decay width for the SM-like Higgs,
and $\Gamma_{h\rightarrow\gamma\gamma}$ is mainly determined by two types of one-loop
Feynman diagrams with either virtual vector bosons or fermions \cite{Gunion},
whose couplings to the Higgs are corrected by factor $(1-2\xi)/\sqrt{1-\xi}$ and $\sqrt{1-\xi}$, respectively.

Apart from the $\gamma$ ray,  DM annihilation into $b\bar{b}$ can also place constraint.
We obtain the cross section  $\langle\sigma_{b\bar{b}} \upsilon_{\rm{rel}}\rangle$  in terms of replacing $\sigma_{\gamma\gamma} \upsilon_{\text{rel}}$ in Eq.(\ref{average}) by \cite{1204.2808}
\begin{eqnarray}
\sigma_{b \bar{b}} v_{\mathrm{rel}} &\approx& \frac{3 m_{b}^{2}\left(1-4 m_{b}^{2} / s\right)^{\frac{3}{2}}}{\pi f^{4}} \left[\frac{1}{4}+\frac{1}{4} \frac{\left(s-\kappa f^{2}\right)^2}{\left(s-m_{h}^{2}\right)^{2}+\Gamma_{h}^{2} m_{h}^{2}} \right.\nonumber \\
&+&\left.\frac{1}{2} \frac{\left(s-m_{h}^{2}\right)\left(s-\kappa f^{2}\right)}{\left(s-m_{h}^{2}\right)^{2}+\Gamma_{h}^{2} m_{h}^{2}}\right],
\end{eqnarray}
where the first, the second and the last term arise from  the contact interaction, the exchange of Higgs and the interference effect, respectively.

Substituting the correlated values of $m_\eta$ and $\kappa$ in Fig.\ref{relic} into Eq.(\ref{average}),
we show in Fig.\ref{gammaray} the numerical results of $\langle\sigma_{\gamma\gamma} \upsilon_{\rm{rel}}\rangle$ and $\langle\sigma_{b\bar{b}} \upsilon_{\rm{rel}}\rangle$ for the representative values of $f$ in Fig.\ref{relic},
where the Fermi-LAT \cite{1506.00013, 1503.02641} and HESS \cite{1301.1173} limits are shown simultaneously.
Compared to the FSDM,
both the values of $\langle\sigma_{\gamma\gamma} \upsilon_{\rm{rel}}\rangle$ and $\langle\sigma_{b\bar{b}} \upsilon_{\rm{rel}}\rangle$ in the case of CSDM are nearly the same in the small $m_\eta$ region.
While the $\gamma$ ray is weak,
the $b\bar{b}$ limit from Fermi-LAT excludes CSDM mass regions $m_{\eta}\leq 47$ GeV and $63$ GeV $\leq m_{\eta}\leq$ $67$ GeV.

\subsection{Precision Test On Higgs Couplings}
The precision measurements on the Higgs couplings are able to effectively constrain the parameter range of $f$.
According to the features of the composite Higgs couplings in Eqs.(\ref{L3})-(\ref{L4}),
we use the conventional two-parametrization fit for our analysis,
under which we have
\begin{eqnarray}{\label{ks}}
k_{V}=\sqrt{1-\xi},~~~~k_{F}=\frac{1-2\xi}{\sqrt{1-\xi}}.
\end{eqnarray}

\begin{figure}
\centering
\includegraphics[width=9.5cm,height=9cm]{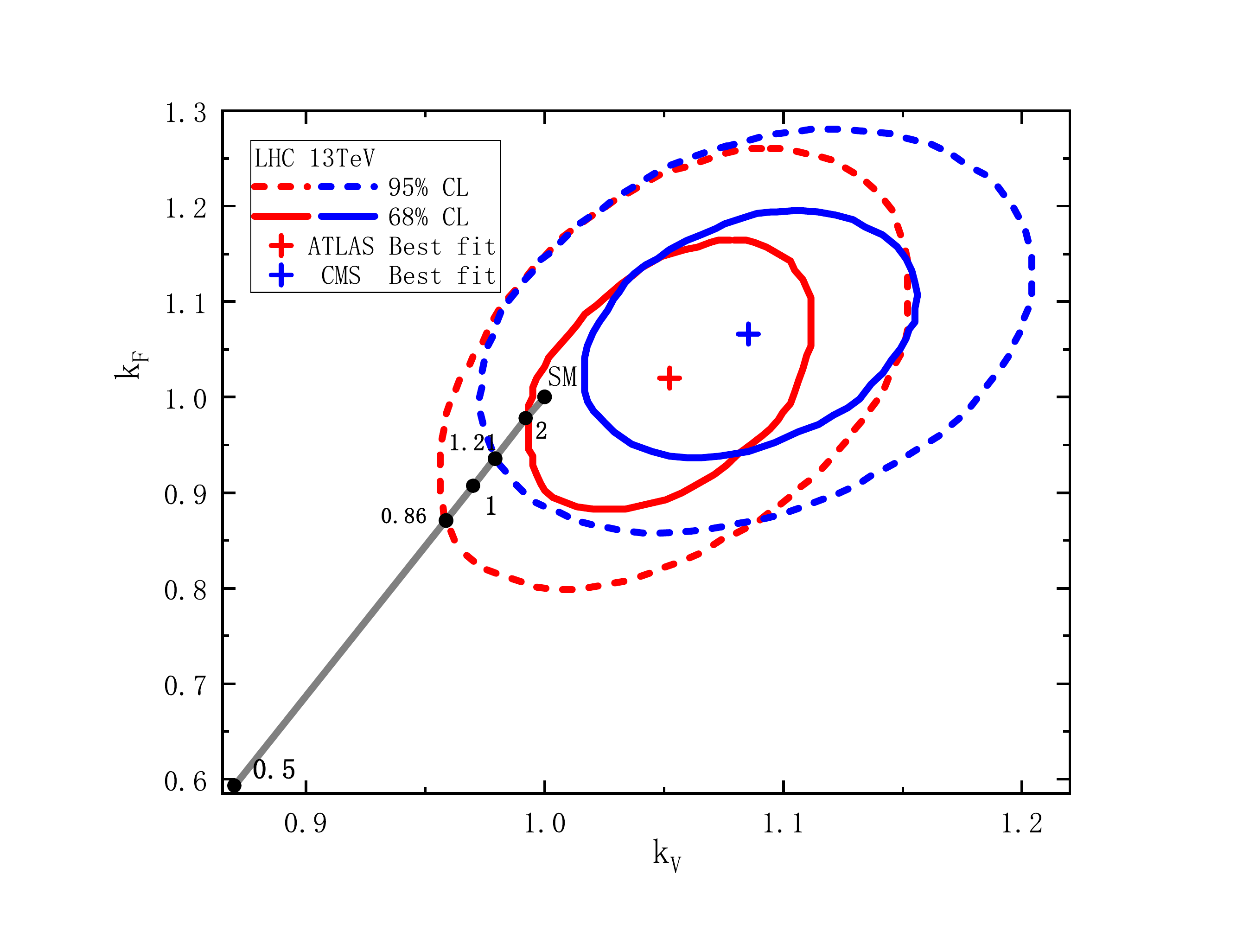}
\centering
\caption{The Higgs couplings for the representative values of $f$ in Fig.\ref{relic},
where both the $68\%$ and $95\%$ contours of the best fits values of $k_F$ and $k_V$ are shown for comparison
and the values of $f$ in unit of TeV at the crossing points are explicitly shown.
We have taken the best fits to these Yukawa coupling constants reported in \cite{Aad:2019mbh, Sirunyan:2018koj}.
 }
\label{Higgsfit}
\end{figure}

Fig.\ref{Higgsfit} shows the constraint on  $f$ from the latest 13-TeV LHC data,
where the ATLAS best fits are given by $k_{V}=1.05$ and $k_{F}=1.05$ \cite{Aad:2019mbh} and the CMS best fits are given by $k_{V}=1.08$ and $k_{F}=1.06$ \cite{Sirunyan:2018koj} respectively.
This figure indicates that
the latest ATLAS and CMS results have excluded the parameter range $f <0.86$ TeV and $f <1.21$ TeV at $95\%$ CL,  respectively.
These lower bounds will be significantly improved at the future LHC, which makes the precision tests on the Higgs couplings more competitive\footnote{Although strong, this indirect constraint may be however evaded in the situation with either non-minimal matter content or non-fundamental representation for the composite fermions.}
than  the precision measurements on the electroweak observables \cite{ALEPH:2005ab}.
In what follows, we will not discuss the case with $f=500$ GeV.

\subsection{Precision Test On Higgs Decay}
\begin{figure}
\centering
\includegraphics[width=9.5cm,height=9cm]{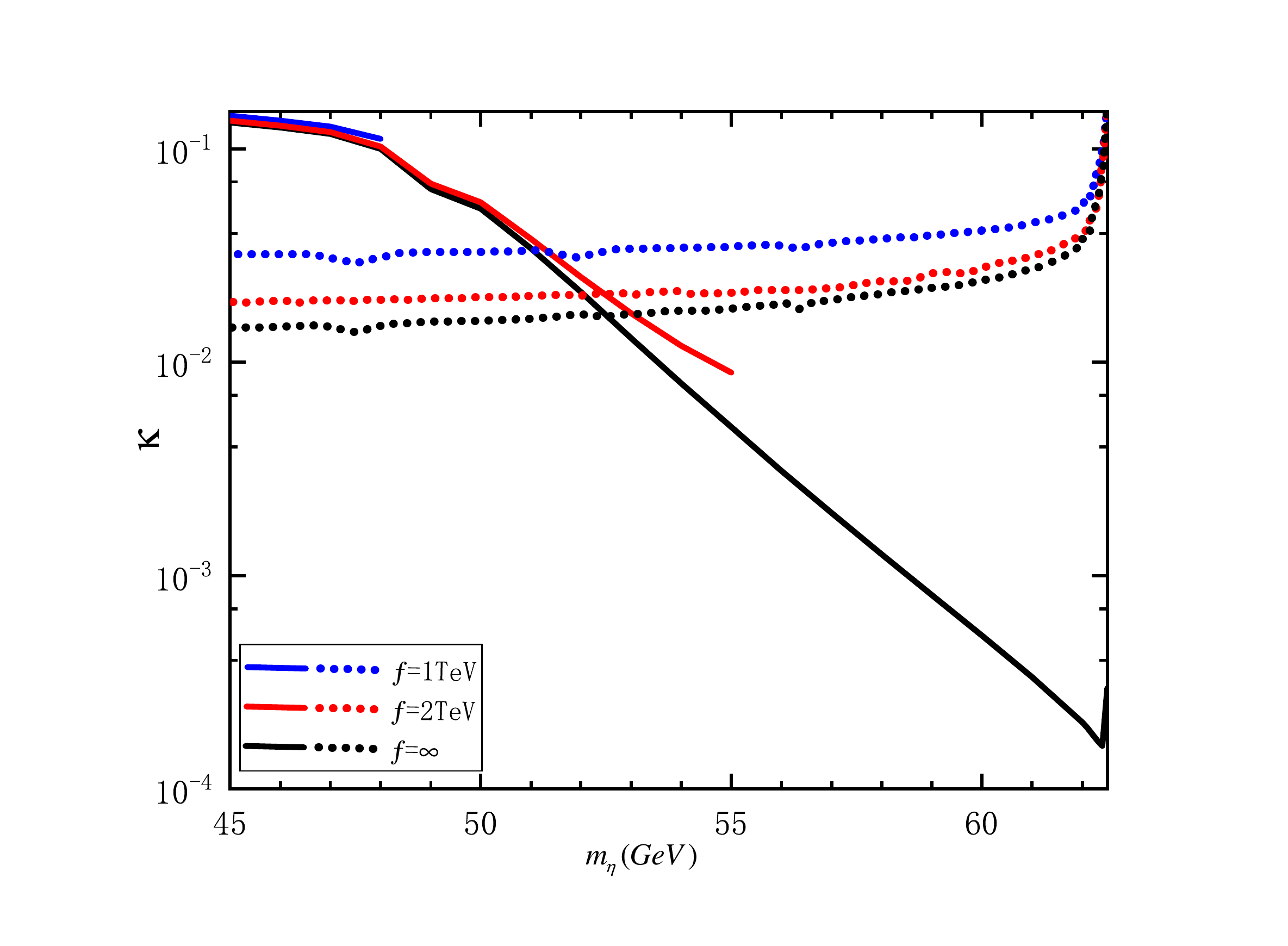}
\centering
\caption{Same as Fig.\ref{relic} with the contours of Higgs invisible decay width $\rm{Br}(h\rightarrow\eta\eta)=16\%$ at 68 $\%$ CL \cite{1606.02266} (in dotted), above which the CSDM mass region is excluded.}
\label{invisible}
\end{figure}

In the CSDM mass region with $m_{\eta}<m_{h}/2$,
the composite Higgs can directly decay into the $\eta$ pair either via the Higgs portal interactions in Eq.(\ref{L2}) or the contact interactions in Eq.(\ref{L3}).
The derivative interactions in Eq.(\ref{L2}) result in a modification to the effective coupling in the Higgs invisible decay,
while the contact interactions contribute to Higgs invisible decay mainly through top quark induced process.
All of the loop effects are controlled by the magnitude of $\xi$.
Without the loop effect, the decay width is approximated as
\begin{eqnarray}{\label{width}}
\Gamma(h\rightarrow \eta\eta)\approx\frac{\upsilon^{2}}{32\pi m_{h}}\left(\kappa-\frac{2m^{2}_{h}}{\upsilon^{2}}\xi \right)^{2}\sqrt{1-\frac{4m^{2}_{\eta}}{m^{2}_{h}}}.
\end{eqnarray}

We show in Fig.\ref{invisible} the contours of the latest experimental bound on the Higgs invisible decay width $\rm{Br}(h\rightarrow\eta\eta)\leq 16\%$ at 68$\%$ CL  \cite{1606.02266} for the representative values of $f$ as in Fig.\ref{relic},
above which the CSDM mass region is excluded.
Compared to the FSDM, this constraint on the CSDM is slightly weaker.
The reason is due to a mild cancellation between the two classes of interactions in Eq.(\ref{width}) given nearly the same $\kappa$ in the mass region $m_{\eta}<m_{h}/2$ regardless of the value of $f$, see Eq.(\ref{relic}).
As a result, the constraint from the Higgs invisible decay is relaxed for finite $f$.
Nevertheless, the absence of the resonant mass region for small $f$ makes this relaxation useless.

The observation holds even with the loop effects taken into account.
For example, the top-loop induced contribution modifies $\kappa$ in Eq.(\ref{width}) by a factor
$\sim \xi\left(\frac{m_{t}}{\upsilon}\right)^{3}\log\left(\frac{m_{t}}{\mu}\right)$, with $\mu$ a cut-off scale.
For $f$ larger than 1 TeV,  it is obviously smaller than $\kappa$.

\section{Direct Detection At LHC}
In this section, we turn to the direct production of the CSDM pair at the LHC.
To calculate the numbers of events of relevant signals and their SM backgrounds,
we use FeynRules \cite{1310.1921} to generate model files prepared for MadGraph5 \cite{1405.0301} that includes Pythia 6 \cite{0603175} for parton showering $\&$ hadronazition, and Delphes 3 \cite{1307.6346} for fast detector simulation.
The leading-order events are obtained in terms of MadGraph5 by extracting samples from the CSDM parameter space in Fig.\ref{relic}.

\begin{figure*}
\centering
\includegraphics[width=8.5cm,height=9cm]{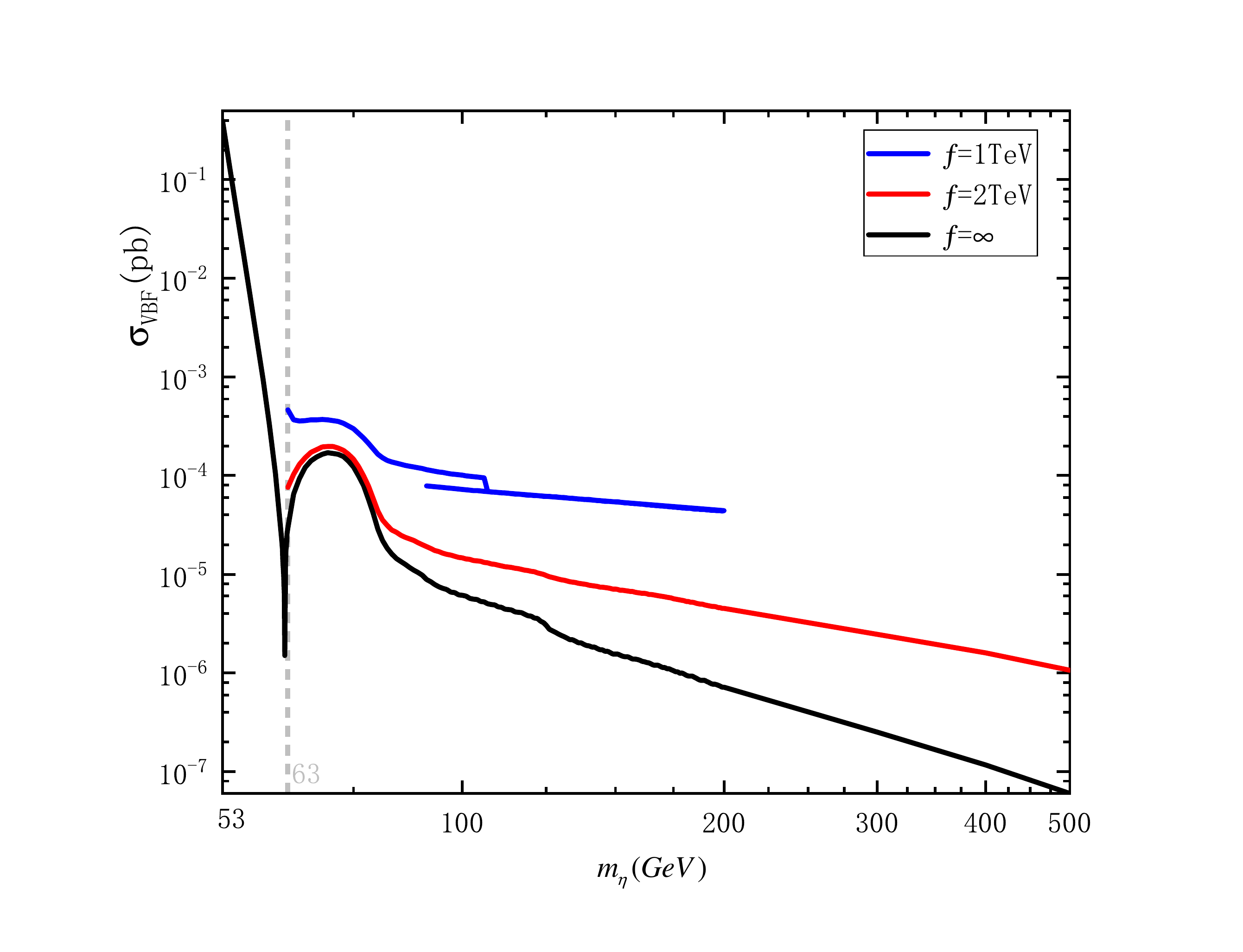}
\includegraphics[width=8.5cm,height=9cm]{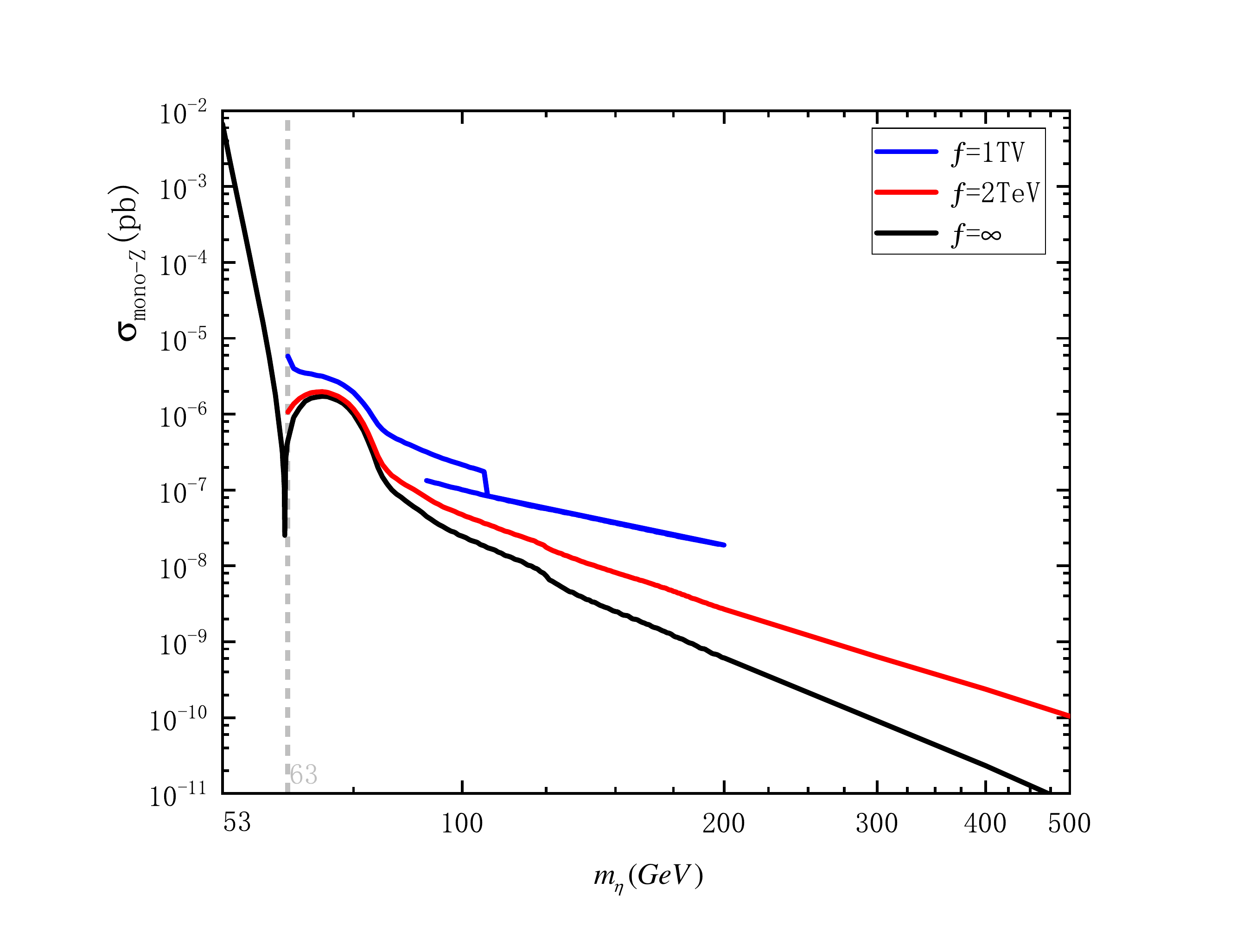}
\centering
\caption{Cross sections of the VBF (left)  and mono-Z (right) process at the 14 TeV LHC for the values of $f$ as in Fig.\ref{relic}, respectively.}
\label{higgsportal}
\end{figure*}

From the Higgs portal in Eq.(\ref{L2}), the $\eta$ pair production at the LHC is similar to that of FSDM.
The discovery channels mainly include the vector boson fusion (VBF) process
\begin{eqnarray}{\label{vbf}}
pp \rightarrow  jj h^{*}\rightarrow jj\eta\eta,
\end{eqnarray}
and the mono-Z process
\begin{eqnarray}{\label{monoz}}
pp \rightarrow Zh^{*} \rightarrow Z\eta\eta,
\end{eqnarray}
where $h$ is virtual for $m_{\eta}>m_{h}/2$,
and the two jets in Eq.(\ref{vbf}) can be either the same or different.
These processes have been used to derive the prospect of the resonant mass region $m_{\eta}\sim m_{h}/2$ at the LHC for the FSDM \cite{1112.3299,1205.3169, Han:2016gyy}.
Unlike the FSDM, the main contribution to the production cross sections of these two signal channels at the 14 TeV LHC
is dominated by the derivative interactions.
Although the derivative interactions enhance the production cross sections, as illustrated in Fig.\ref{higgsportal},
compared to the SM  cross sections
of $54$ pb, $9.6$ pb and $30.9$ pb for $W+$jets, $Z+$jets and mono-Z respectively,
they are about at least five orders of magnitude smaller.
So large gaps between the cross sections make these processes unlikely to constrain the CSDM at the HL-LHC with an integrated luminosity  3000 $\rm{fb}^{-1}$.
We draw this conclusion based on the 13-TeV CMS cuts reported in \cite{Sirunyan:2018owy}  and \cite{Sirunyan:2017onm} for the VBF and mono-Z respectively.

In addition,
the contact interactions in Eq.(\ref{L3}) provide alternative production processes different from those of FSDM.
Among them, the top-loop induced gluon gluon fusion (GGF) process\footnote{Concretely speaking,
both the contact and Higgs interactions contribute to this GGF process,
with the former dominating over the later.}
\begin{eqnarray}{\label{ggf}}
pp\rightarrow jj\eta\eta,
\end{eqnarray}
has the largest signal rate.
Besides the GGF process, there are also signal channels with top quark pair such as $pp\rightarrow \bar{t}t \eta\eta \rightarrow \bar{b}b j j j j \eta\eta$ with hadronic final states and $pp\rightarrow \bar{t}t \eta\eta \rightarrow \bar{b}b j j \eta\eta \ell \nu$ with leptonic final state(s) \cite{ATLAS:2020llc},
whose SM backgrounds are mainly given by $pp\rightarrow \bar{b}b j j j j\nu\nu$ and $pp\rightarrow \bar{b}b j j \ell \nu$, respectively.
The GGF process has the cross section of order up to $\sim 10^{2}$ fb,
while the processes with the top quark pairs have cross section of order up to $\sim 10^{-1}$ fb.
Unfortunately, all of these production cross sections are too small.
Take the GGF process for example.
Compared to its SM background with the cross section of order $\sim 6\times 10^4$ pb,
the GGF process fails to provide any useful constraint,
no matter how the selection of events are performed.

\begin{figure*}
\centering
\includegraphics[width=16cm,height=20cm]{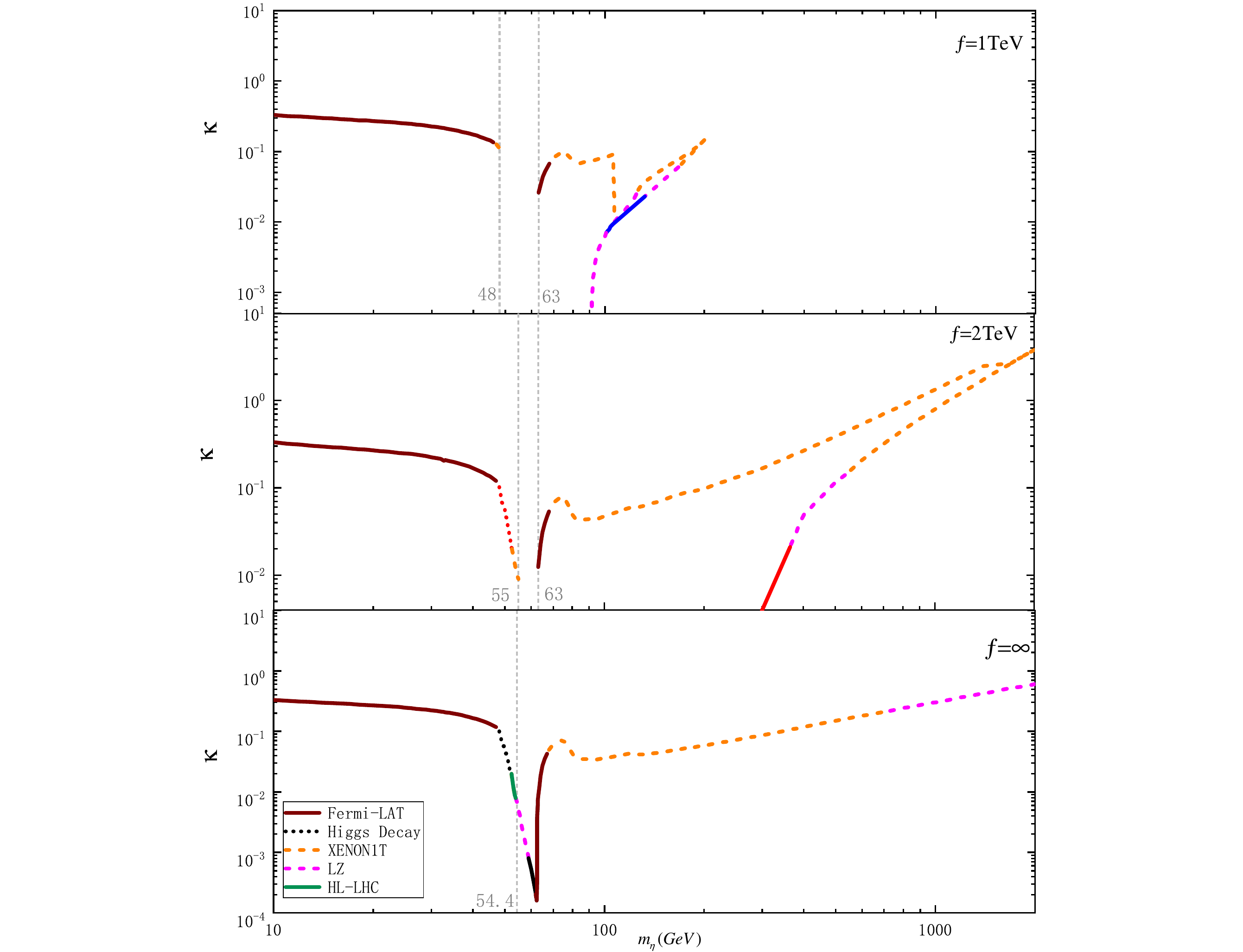}
\centering
\caption{The CSDM mass subject to the combination of direct detection (current XENON1T and future LZ limits) as well as the indirect constraints from the Fermi-LAT limits on the DM annihilation cross sections,
the Higgs invisible decay and the precision tests on the Higgs couplings,
where  the conservative ATLAS bound $f>0.86$ TeV at 95$\%$ CL has been taken.
The FSDM (the lowest plot) is shown for comparison, where the 5$\sigma$ discovery limit \cite{Han:2016gyy} at the HL-LHC is highlighted in dark green.
The references of the other colors are the same as before.}
\label{total}
\end{figure*}

Based on the null results from the VBF, mono-Z and GGF processes,
the minimal CSDM with mass $m_{\eta}>m_{h}/2$ is
totally invisible at the high-luminosity(HL)-LHC with the integrated luminosity $3000$ fb$^{-1}$.
Consider that the CSDM couplings to the SM Higgs and fermions aren't obviously altered in the situation of non-minimal scenarios,
we infer that the small signal rate of CSDM at the LHC in the large DM mass region is probably a general result.

\section{Conclusions}
In this study we have made a comprehensive investigation on the CSDM,
based on the framework of parametrization which can help us estimate the general phenomenological status.
Although totally different from the FSDM,
the CSDM mimics the FSDM when the scale of global symmetry breaking $f$ is far than the weak scale.
But their differences become ``visible" as $f$ decreases to the order of TeV scale (where the fine tuning is small).
The minimal CSDM has been exposed by imposing both direct and indirect constraints.
Fig.\ref{total} shows how to differentiate it from the FSDM as what follows.
\begin{itemize}
\item  Disappearing resonant mass region.  Due to the derivative interaction the resonant mass region gradually disappears from $f=\infty$ to $f=1$ TeV in Fig.\ref{total}.
\item  Small SI DM-nucleon scattering cross section in certain mass region.
Instead of the exclusion mass bound larger than $\sim 700$ GeV in the FSDM,
a large part of the CSDM mass window between $\sim 67$ GeV and $\sim 600$ GeV is still beneath the current XENON1T limit.
Since future LZ experiment can reach a partial of this mass region,
they are very useful in distinguishing the CSDM from the FSDM.
\item The absence of CSDM at the HL-LHC.
Compared to certain signal reach near the resonant region in the FSDM as shown by the dark green curve in Fig.\ref{total},
the disappearing resonant mass region together with the small signal rates in the larger DM mass region for the CSDM make the HL-LHC an alternative platform to distinguish these two DM models.
\end{itemize}

\section*{Acknowledgments}
Zheng would like to thank the Department of Physics at Harvard University for hospitality, where this work was initiated.The research is supported in part by the National Natural Science Foundation of China with Grant No. 11775039 and the Fundamental Research Funds for the Central Universities at Chongqing University with Grant No. cqu2017hbrc1B05. S.X. is supported by Zhoukou Normal University under Grant No. ZKNUC2021006.

\end{document}